# SECONDARY ELECTRON EMISSION YIELD IN THE LIMIT OF LOW ELECTRON ENERGY

A. N. Andronov, A. S. Smirnov, *St. Petersburg State Polytechnic University, St. Petersburg, 195251 Russia*
I. D. Kaganovich, E. A. Startsev, and Y. Raitses, *Princeton Plasma Physics Laboratory, Princeton, NJ 08543, USA*
V. I. Demidov, *West Virginia University, USA*

*Abstract*

Secondary electron emission (SEE) from solids plays an important role in many areas of science and technology.[1] In recent years, there has been renewed interest in the experimental and theoretical studies of SEE. A recent study proposed that the reflectivity of very low energy electrons from solid surface approaches unity in the limit of zero electron energy[2,3,4]. If this was indeed the case, this effect would have profound implications on the formation of electron clouds in particle accelerators,[2-4] plasma measurements with electrostatic Langmuir probes, and operation of Hall plasma thrusters for spacecraft propulsion[5,6]. It appears that, the proposed high electron reflectivity at low electron energies contradicts to numerous previous experimental studies of the secondary electron emission[7]. The goal of this note is to discuss possible causes of these contradictions.

## THEORETICAL DESCRIPTION OF SECONDARY ELECTRON EMISSION IN THE LIMIT OF LOW ENERGY

Authors of the Refs. [2-4] suggest that the theoretical description of the elastic backscattering of low-energy electrons at solid surface could be adequately described by a very simplified one-dimensional model of the quantum reflection of electron plane waves incident on an abrupt potential step of height, $eV_i$ of the internal potential. For example, in R. Cimino's work[3] it was assumed that $eV_i$ =150 eV, which is unrealistically large number for typical internal potential of order several electronvolts. Such a large number was assumed to explain experimental result that the secondary electron emission yield (SEEY) tends to unity as energy approaches 0 at electron energies below 10eV. It was assumed that electrons are reflected back only at the surface potential barrier and do not penetrate into the solid. Such description is oversimplified for several reasons.

First, electron interaction with the real surface of a solid target cannot be described by a potential barrier with a sharp step. At distances $x$ greater than the interatomic distance in solids when the metal surface can be considered as perfectly smooth and perfectly conducting, electron experiences an image force, which is equal to $e^2/4x$ (the so-called Schottky effect[8]). At shorter distances the metal surface cannot be considered as perfectly smooth and the work function is determined by the dipole moments of surface atoms, preventing the exit of electrons into vacuum. In this region the potential is a nearly-constant and equal to the inner potential of the solid.

Quantum-mechanical reflectivity, $R$, of slow electrons at the barrier of such a form is taken into account in the Richardson law for the thermionic emission current density[9], $J$: $J=A_0T^2D$ exp $(-\phi/kT)$, where $A_0$ is the Sommerfeld constant=120.4A/cm$^2$K$^2$, $D=1-R$ is transparency of the barrier, and $\phi$ is the local work function of the sample. It is shown experimentally that for thermal electrons (energy in vacuum less than 0.1 eV) the average coefficient of reflection at the surface is less than 10 percent, and in most cases much less. According to the laws of quantum mechanics, the reflection does not depend on the direction of electron velocity. Therefore, coefficient of reflection at the barrier must be the same for both, the thermal electrons that are escaping from the solid, and for the incident primaries.

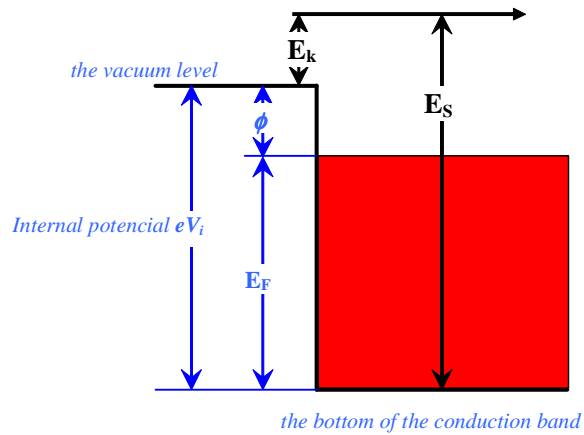

Figure 1: The energy level diagram.

Additional contribution to the elastic reflection is caused by electrons scattered inside a solid target. Incident primary electrons are accelerated by the surface potential barrier to the value $E_s=E_k+E_F+\phi$ (see Fig.1) and then they penetrate into a solid target to a depth of 50-100 Å [10,11]. In that process electrons lose their energy and produce secondary electrons, and are also elastically scattered by the atoms of a solid. This scattering (so called s-scattering) is isotropic and its cross section is determined by the atomic number of scattering atoms and does not depend on energy when the electron wavelength is much greater than inter-atomic distance[12]. A half of the elastically scattered electrons are scattered by an angle of more than π/2 (back-scattered electrons). These electrons also undergo the elastic and inelastic scattering on their way back to the surface and only a small fraction reaches the surface with an initial energy $E_s \geq eV_i$. When an electron passes through the surface potential barrier in vacuum only, the normal to the surface component of the electron momentum changes and the parallel to the surface component of the electron momentum does not change. Therefore, if electron has small kinetic energy in vacuum, $E_k$, this electron cannot overcome the potential barrier after scattering on atoms unless it scattered exactly backward 180 degrees. In the limit $E_k \to 0$, the contribution of these back scattered electrons to the total elastic reflection tends to zero (not to unity). Thus, this process also does not increase the elastic reflection to full 100 percent eflection with kinetic energy of the electrons in vacuum decreasing to zero.

## EXPERIEMTNAL CONSIDERATIONS AND DIFFICULTIES OF MEASUREING THE SECONDARY ELECTRON EMISSION YIELD IN THE LIMIT OF LOW ENERGY

Experimental measurements at low incident electron energy, below 2eV, are extremely challenging. It is very difficult to produce collimated aligned electron beam at such low energy. References 7 and 18 measured reflection coefficient down to 3 eV making use of low-energy electron beam. They observed reflection coefficient below 10 percent for clean targets and 40% for contaminated targets with some absorbed gas on surface. Because it is very difficult to produce electron beam with such low energy, common approach to study secondary electron yield is to use an electron gun at a given fixed energy and slow down electrons by applying retarding potential to the target. But "when the incident energy is decreased by an increase of the negative bias of the sample holder instead of a decrease of the nominal beam energy E, such a bias leads to constant potential surfaces in the vacuum gap on which the incident electrons may be totally reflected and then collected without any contact with the sample surface". Making use of such approach authors of Ref. [13] reported the reflection coefficient 40% whereas authors of Ref. [14] reported reflection coefficient 100%.

The most recent detailed review of the latest research in the field of reflectivity of very low energy electrons from solid surfaces Ref. [1] refers to the Ref. [2] as the main experimental evidence of the proposed hypothesis. The examination of description of experiment conducted by these authors shows that they used experimental setup that is commonly used to measure a contact potential difference between cathode and a target making use of an electron beam (also so-called the Anderson method [15,16,17]). In this method, the contact potential difference between cathode and a target is determined by the condition when the current between cathode and anode (target) is zero. Therefore, in order to achieve that the electron beam with low energy reaches the target, the voltage on the target should be carefully chosen to be equal the beam energy minus the contact potential difference between the cathode and the target. Without taken into account the effect of the contact potential difference between the cathode and the target on beam energy, the incident beam electron may reflect from the retarding potential in the vacuum gap without reaching the target. This reflection from the retarding potential in the vacuum gap then will be inaccurately interpreted as 100% elastic reflection from the surface.

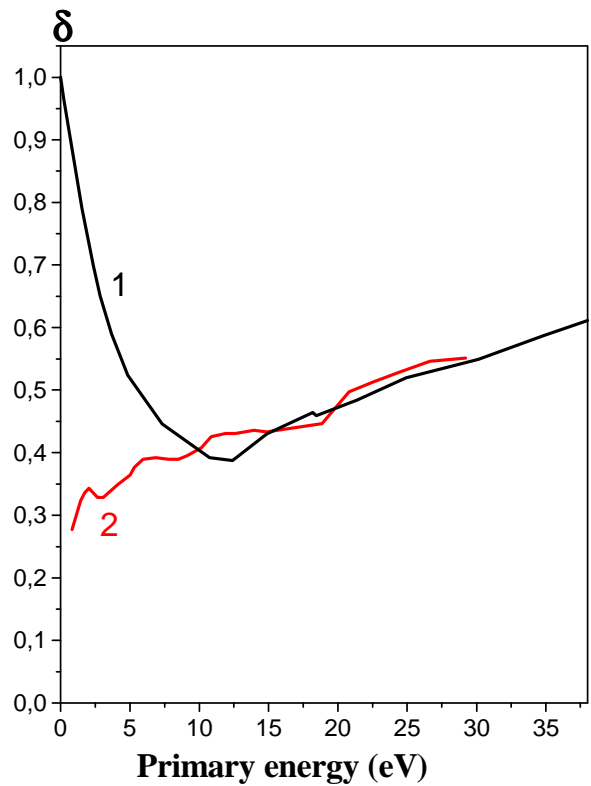

Figure 2: Total secondary electron emission yield of Cu at low electron primary energy $E_p$.
1. Initial part of Fig. 2 taken from the letter [3] for fully scrubbed Cu ($T$=10 K).
2. Experimental data for Cu from Ref. [18] after applying heat in vacuum (taken at room temperature).

It is difficult to access what was the exact reason for artificial 100% reflection reported in the experiments of Ref.[2]. However, comparing the data of Ref.[2] with the data confidently obtained in numerous previous careful measurements reported in Refs.[19,20,21] shows big differences for electron energies in the range 5-10 eV. For example, Fig.2 depicts comparison of the results under discussion with previously published data for copper targets. From Fig.2 it is evident that there is sufficiently good agreement between two data sets for $E_p$>10 eV, but there are drastic differences for $E_p$<10 eV. However, the data of Ref. [3] were taken for a cryogenically cooled target whereas data of Ref. [18] were taken at normal conditions. As follows from the theoretical description that should not make a big difference.

In summary, we presented ample evidence that numerous previous measurements in the low energy range shows reflection coefficient of about 7% in the range of few electron volts. We also presented compelling arguments that Refs. [2-4] provide contradictory account of SEE at low energies when compared with other numerous previous measurements in the low energy range given by authors of Refs. [7, 18-21]. In addition, there are straightforward theoretical considerations that support the claim that the reflection coefficient should remain small even in the limit of very low electron energy.

## ACKNOWLEDGMENT

This work was supported by the U.S. Department of Energy and the U.S. Air Force Office of Scientific Research.